\documentclass[twocolumn,showpacs,preprintnumbers,amsmath,amssymb,prl]{revtex4}

\usepackage{graphicx}
\usepackage{dcolumn}
\usepackage{bm}
\usepackage{tabularx}
\usepackage{amsmath}

\begin{document}

\title{Ballistic magnon transport and phonon scattering in the antiferromagnet Nd$_2$CuO$_4$}

\author{S. Y. Li$^1$, Louis Taillefer$^{1*}$, C. H. Wang$^2$, and X. H. Chen$^2$}

\affiliation{$^1$D{\'e}partement de physique, Regroupement
Qu{\'e}b{\'e}cois sur les Mat{\'e}riaux de Pointe, Universit{\'e}
de Sherbrooke, Sherbrooke, Canada\\
$^2$Hefei National Laboratory for Physical Science at Microscale
and Department of Physics, University of Science and Technology of
China, Hefei, Anhui 230026, P. R. China}

\date{\today}

\begin{abstract}
The thermal conductivity of the antiferromagnet Nd$_2$CuO$_4$ was
measured down to 50 mK. Using the spin-flop transition to switch
on and off the acoustic Nd magnons, we can reliably separate the
magnon and phonon contributions to heat transport. We find that
magnons travel ballistically below 0.5 K, with a thermal
conductivity growing as $T^3$, from which we extract their
velocity. We show that the rate of scattering of acoustic magnons
by phonons grows as $T^3$, and the scattering of phonons by
magnons peaks at twice the average Nd magnon frequency.
\end{abstract}


\maketitle

The collective excitations of an antiferromagnet are magnons. Like
electrons and phonons, magnons can transport heat. Several groups
have recently used heat transport to probe spin excitations in
quantum magnets \cite{Lorenz,Sales,Hess,Sologubenko,Hess2,Takeya}.
The results tend to be difficult to interpret because magnons
interact strongly with phonons. However, as $T \rightarrow 0$, the
two become decoupled and their respective mean free paths are
expected to reach the size of the sample. This ballistic regime is
easily reached for phonons, but it has rarely been demonstrated
for magnons, because acoustic magnons are almost always gapped. If
they were not gapped, their thermal conductivity would be hard to
separate from that of acoustic phonons, as both are expected to
exhibit a $T^3$ temperature dependence in the ballistic regime. In
this context, the antiferromagnetic insulator Nd$_2$CuO$_4$ offers
a rare opportunity. Jin {\it et al.} have recently shown that in
this material magnon heat transport can be switched on and off
with a magnetic field via a spin-flop transition \cite{Jin}.

In this Letter, we use this ``switch'' to investigate the physics
of magnons. First, by cooling down to 50 mK, we show that the
ballistic regime for magnon transport, characterized by a clear
$T^3$ conductivity, sets in below 0.5 K. The magnitude gives us
the magnon velocity, whose value agrees with a recent calculation
\cite{Sachidanandam}. Secondly, we show how above 0.5 K magnons
are scattered by phonons at a rate proportional to $T^3$. Finally,
we show that the scattering of phonons by magnons is strongly
peaked at a temperature where the optimum phonon energy equals
twice the average magnon energy, in line with a model of dominant
one phonon-two magnon processes \cite{Dixon}.

Nd$_2$CuO$_4$ is the insulating parent compound of the
electron-doped high-$T_c$ superconductor. The spins on the
Cu$^{2+}$ ions order antiferromagnetically below $T_N \approx$ 275
K, with spins lying in the CuO$_2$ planes. As the temperature is
lowered, the ordered spin structure undergoes a series of
re-arrangements, to reach a non-collinear state at low temperature
($T <$ 30 K), with the spin direction changing by 90 degrees in
alternate planes along the $c$-axis
\cite{Skanthakumar2,Sachidanandam,Petitgrand,Li}. Due to the
magnetic exchange interaction between Cu$^{2+}$ and Nd$^{3+}$,
moments on the Nd site also order in the same structure. The
magnon spectrum of Nd$_2$CuO$_4$, studied by neutron scattering
\cite{Ivanov,Petitgrand,Henggeler,Casalta}, is characterized by
two sets of branches, associated respectively with spins on Cu and
Nd sites. Cu magnons have energies above 5 meV
\cite{Ivanov,Petitgrand}, and four optical Nd magnon branches lie
in the range 0.2 to 0.8 meV \cite{Henggeler,Casalta}. A recent
model predicts that the Nd spin sublattice should also support an
{\it acoustic} magnon mode, with a slope of 10 meV \AA\ and a very
small gap (5 $\mu$eV) \cite{Sachidanandam}. An in-plane magnetic
field of 4.5 T causes a spin-flop transition from non-collinear to
collinear state. This change of spin structure modifies the Nd
magnon spectrum, as evidenced by the additional heat conduction
which appears at the transition \cite{Jin}.

Single crystals of Nd$_2$CuO$_4$ were grown by a standard flux
method. Our sample was annealed in helium for 10 h at 900 $^o$C,
and cut to a rectangular shape of dimensions 1.50 $\times$ 0.77
$\times$ 0.060 mm$^3$ (length $\times$ width $\times$ thickness in
the $c$ direction). Contacts were made with silver epoxy, diffused
at 500 $^o$C for 1 hour. The thermal conductivity $\kappa(T)$ was
measured using a standard steady-state technique \cite{Mike}, in a
dilution refrigerator (50 mK to 2 K, with RuO$_2$ thermometers),
and in a $^4$He cryostat (above 2 K, with Cernox thermometers).
The magnetic field $H$ was applied perpendicular to the heat
current, both in-plane and along the $c$-axis. Note that in a
magnetic insulator, the thermal conductivity is the sum of phonon
and magnon conductivities: $\kappa(T)$ = $\kappa_p(T)$ +
$\kappa_m(T)$.

{\it Ballistic regime.} ---  Fig. 1 shows the temperature
dependence of $\kappa$, in $H$ = 0, 10 T $\perp c$, and 10 T //
$c$ \cite{note}. The in-plane magnetic field causes a huge
enhancement of $\kappa$, as discovered by Jin {\it et al.} above 2
K \cite{Jin}. The advantage of going lower in temperature by a
factor 40 is that one can then unambiguously decouple $\kappa_p$
and $\kappa_m$. In Fig. 2 we plot the zero-field data (upper
panel), as $\kappa(0)$ vs $T^{2.6}$, and the difference between
in-field data and zero-field data (lower panel), as $\Delta\kappa$
= $\kappa$(10 T $\perp c) - \kappa(0)$ vs $T^3$. In both cases,
the data follows a straight line below about 0.5 K, which defines
the ballistic regime for both types of carriers, wherein the mean
free path is limited by sample boundaries.

\begin{figure}[t]
\centering \resizebox{\columnwidth
}{!}{\includegraphics{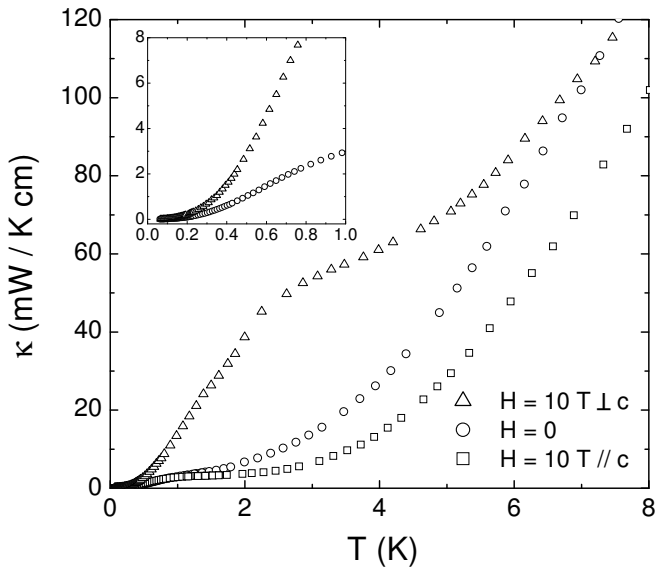}} \caption{\label{fig1} Temperature
dependence of the in-plane thermal conductivity $\kappa$ of
Nd$_2$CuO$_4$, in zero magnetic field (circles; $H$ = 0) and in a
field of 10 T applied either in the basal plane (triangles; $H
\perp c$) or along the $c$-axis (squares; $H // c$). {\it Inset}:
Zoom below 1 K.}
\end{figure}

a) {\underline {Phonons ($T <$ 0.5 K)}}. As $T \rightarrow 0$, the
phonon mean free path in insulators becomes limited only by the
physical dimensions of the sample. In this ballistic regime, the
phonon conductivity is given by \cite{Ziman}:
\begin{equation}
 \kappa_p = \frac{2}{15} \pi^2k_B(k_BT/\hbar)^3 <\hspace{-3pt}v_p^{-2}\hspace{-3pt}>
 l_0\label{eq:kappa_ph},
\end{equation}
where $<\hspace{-3pt}v_p^{-2}\hspace{-3pt}>$ is the inverse square
of the sound velocity averaged over the three acoustic branches in
all {\it {\bf q}} directions, and $l_0$ is the
temperature-independent mean free path, given by the
cross-sectional area $A$ of the sample: $l_0$ = 2$\sqrt{A/\pi}$.
For our sample, $l_0$ = 0.24 mm.

For samples with smooth surfaces, a fraction of phonons undergo
specular reflection and the mean free path exceeds $l_0$,
estimated for diffuse scattering. Because specular reflection
depends on the ratio of phonon wavelength to surface roughness,
the mean free path depends on temperature and $\kappa_p$ no longer
grows strictly as $T^3$ \cite{Ziman}. The general result is that
$\kappa_p$ $\propto$ $T^{\alpha}$, with $\alpha < 3$ \cite{Pohl}.
Typical values of $\alpha$ are 2.6-2.8 (see \cite{Mike} and
references therein). For instance, $\kappa_p$ $\propto$ $T^{2.7}$
in the insulator La$_2$CuO$_4$ \cite{Dave} (where all magnons are
gapped). In Fig. 2, the zero-field conductivity of our
Nd$_2$CuO$_4$ crystal can be fitted to $\kappa_p = b T^{\alpha}$
below 0.4 K, with $\alpha$ = 2.6. This is a first indication that
$\kappa(0)$ is mostly due to phonons, with a ballistic regime
below $\sim$ 0.5 K.

The second argument comes from the magnitude of $b$. Taking $l_0$
= 0.24 mm as a lower bound on the mean free path and applying Eq.
(1) at $T$ = 0.4 K, where $\kappa(0)$ = 0.61 mW / K cm, we obtain:
$\bar{v}_p$ = ($<\hspace{-3pt}v_p^{-2}\hspace{-3pt}>$)$^{-1/2}$ =
3.2 $\times$ 10$^3$ m/s. A mean free path larger by a factor of 2
would give $\bar{v}_p$ = 4.5 $\times$ 10$^3$ m/s. These values are
in good agreement with the measured sound velocities of
Nd$_2$CuO$_4$: $v_L$ = 6.05 $\times$ 10$^3$ m/s and $v_T$ = 2.46
$\times$ 10$^3$ m/s for longitudinal and transverse modes along
[100] \cite{Fil}. So it appears that phonons account for the full
$\kappa(0)$. In a third argument presented later, we make the case
that magnon transport is negligible in the non-collinear state at
{\it any} $T$, so that $\kappa(0) \simeq \kappa_p$ at all $T$.

\begin{figure}[t]
\centering \resizebox{\columnwidth
}{!}{\includegraphics{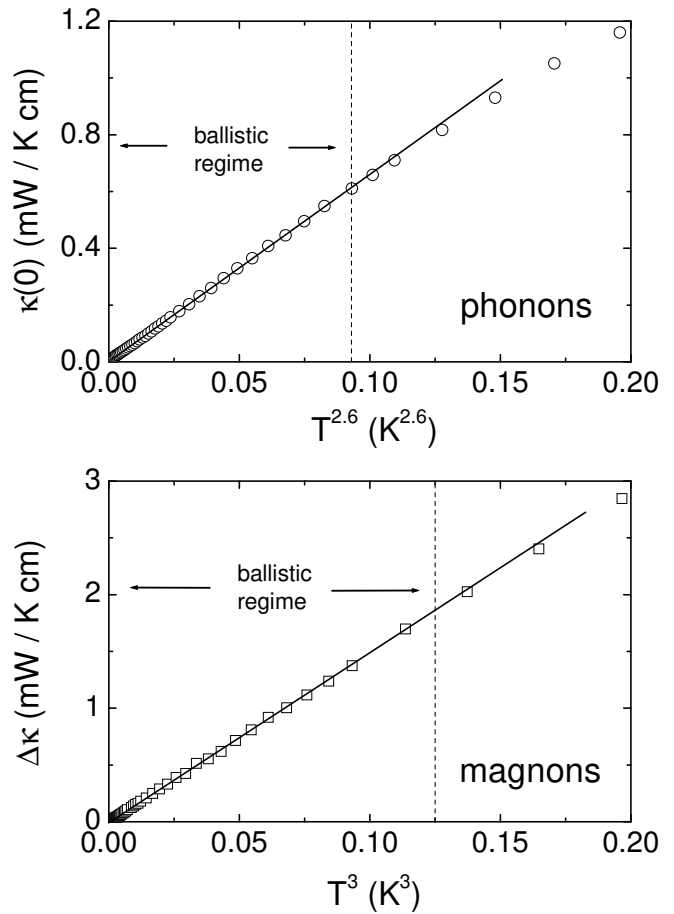}} \caption{\label{fig2} {\it Top
panel}: Zero-field thermal conductivity $\kappa(0)$ vs $T^{2.6}$.
The heat at $H$ = 0 is carried entirely by phonons: $\kappa_p$
$\simeq$ $\kappa$(0) (see text). The line is a linear fit below
0.4 K. {\it Bottom panel}: The additional thermal conductivity
induced by a 10 T in-plane field, $\Delta\kappa$ = $\kappa$(10 T
$\perp c$) - $\kappa$(0), plotted as $\Delta\kappa$ vs $T^3$. The
line is a linear fit below 0.5 K. The additional heat is carried
entirely by acoustic magnons: $\kappa_m$ = $\Delta\kappa$ (see
text).}
\end{figure}

b) {\underline {Magnons ($T <$ 0.5 K)}}. Magnon transport is
revealed by looking at the difference between the two curves in
the inset of Fig. 1, $\Delta\kappa$ = $\kappa$(10 T $\perp c)$ -
$\kappa(0)$, plotted in the bottom panel of Fig. 2. We get the
striking result that $\Delta\kappa$ = $dT^3$, below 0.5 K. Such a
field-induced additional $T^3$ contribution to $\kappa$ at low $T$
can be attributed unambiguously to acoustic magnons. To our
knowledge, this is the first observation of the characteristic
$T^3$ dependence of magnon thermal conductivity \cite{Note}. As
bosons with a linear energy dispersion at low $q$, acoustic
magnons should contribute to heat transport in the same way as
acoustic phonons, provided there is no gap in their spectrum.
Hence for gapless acoustic magnons, Eq. 1 should apply in the
ballistic regime, with $<\hspace{-3pt}v_m^{-2}\hspace{-3pt}>$
replacing $<\hspace{-3pt}v_p^{-2}\hspace{-3pt}>$ \cite{vp}. Note,
however, that the prefactor will depend on the number of magnon
branches. In the simplest case, one expects two degenerate magnon
branches, with the degeneracy lifted by a magnetic field. As our
data is in 10 T, we assume that only one branch lies at zero
energy. Hence:
\begin{equation}
 \kappa_m = \frac{1}{3} \times \frac{2}{15} \pi^2k_B(k_BT/\hbar)^3 <\hspace{-3pt}v_m^{-2}\hspace{-3pt}>
 l_0\label{eq:kappa_ph}.
\end{equation}
Assuming that $\kappa(0)$ = $\kappa_p$ and that $\kappa_p$ is
independent of magnetic field in the ballistic regime, we have
$\kappa_m$ = $\kappa$ - $\kappa_p$ = $\Delta\kappa$, so that
$\kappa_m$ = $dT^3$, with $d$ = 15 mW K$^{-4}$ cm$^{-1}$. Using
Eq. 2 with $l_0$ = 0.24 mm, we get $\bar{v}_m$ =
($<\hspace{-3pt}v_m^{-2}\hspace{-3pt}>$)$^{-1/2}$ = 1.5 $\times$
10$^3$ m/s.

With the important caveat that their calculation was done for the
{\it non-collinear} state ($H$ = 0), our findings agree with the
theoretical calculations of Sachidanandam {\it et al.}
\cite{Sachidanandam}: 1) the acoustic mode they predicted is
confirmed, 2) the predicted magnitude of the gap (5 $\mu$eV) is
consistent with our lowest temperature (50 mK), and 3) the mode
dispersion (or magnon velocity) is of the right magnitude (10 meV
\AA\ = 1500 m/s).

An intriguing difference between phonons and magnons in the
ballistic regime is that only the former appear to undergo
specular reflection. Indeed, the perfect $T^3$ dependence of
$\kappa_m$ implies that the fraction of magnons that are
specularly reflected must be negligible, while it is not for
phonons. We speculate that the strong anisotropy expected of the
magnon velocity in Nd$_2$CuO$_4$, compared to the roughly
isotropic phonon velocity, might account for this difference. It
is also likely that the spin structure is disordered at the
surface so that the surface is rougher magnetically than
structurally.

{\it Inelastic regime}. --- Beyond the ballistic regime, the
temperature interval up to 10 K or so is dominated by
phonon-magnon scattering: a) acoustic phonons are scattered
strongly by magnons, and b) acoustic magnons are scattered
strongly by phonons.

\begin{figure}[t]
\centering \resizebox{\columnwidth
}{!}{\includegraphics{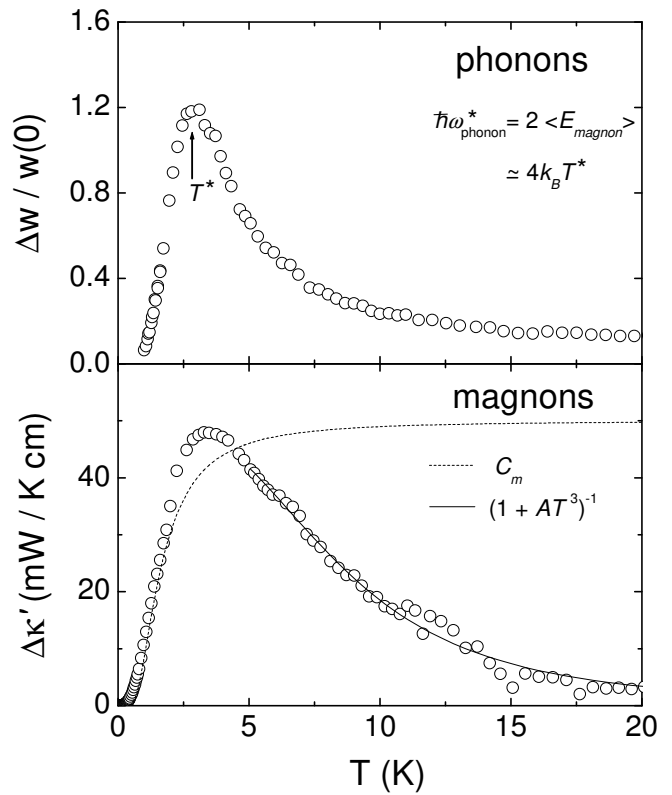}} \caption{\label{fig3} {\it Top
panel}: Normalized thermal magneto-resistance vs temperature, for
$H // c$. This is the field-dependent thermal resistivity of
phonons. The arrow indicates the temperature at which magnon
scattering is maximum: $T^{\star}=3$ K. $T^{\star}$ is directly
related to the magnon energy $<E_{magnon}> $ via the peak phonon
frequency, $\omega_{phonon}^{\star}$ (see text). {\it Bottom
panel}: Difference in thermal conductivity caused by rotating a 10
T field from in-plane to $c$-axis, $\Delta\kappa$' = $\kappa$(10 T
$\perp c$) - $\kappa$(10 T // $c$), vs temperature. This is the
conductivity of magnons. The dotted line is the estimated magnon
specific heat (see text) and the solid line is a fit as
indicated.}
\end{figure}

a) {\underline {Phonons ($T >$ 0.5 K)}}. A 10 T field along the
$c$-axis leaves the system in its non-collinear state and as seen
in Fig. 1, it has negligible impact below 1 K. Above 1 K, however,
it clearly suppresses $\kappa$, relative to $\kappa(0)$. The 10 T
// $c$ curve is striking: the rapid growth in phonon conductivity
generally observed in insulators is almost entirely stopped
between 1 and 3 K. An extremely effective inelastic scattering
mechanism switches on near 1 K and off above 3 K. This range of
energies is precisely that of Nd magnons in Nd$_2$CuO$_4$.

The theory of phonons scattered by magnons was given a thorough
treatment by Dixon \cite{Dixon}, who argues that the one
phonon-two magnon process is dominant. Dixon gets two main
results: 1) the scattering peaks at a temperature such that the
optimum phonon frequency ($\sim$ 4$k_BT$) is equal to the average
energy of two magnons, and 2) a magnetic field enhances the
effectiveness of magnon scattering, especially when the field is
perpendicular to the spin direction. We attribute the suppression
of $\kappa$ induced by applying $H$//$c$ to this enhanced
effectiveness.

Phonons are scattered by magnons and several other processes
(boundaries, defects, other phonons). However, because only magnon
scattering depends on magnetic field, we can bring out the role of
magnon scattering by focusing on the {\it change} in thermal {\it
resistivity}, $w = 1/\kappa$, {\it i.e.} the thermal
magnetoresistance $\Delta w \equiv w$(10 T // $c$) - $w$(0). In
the top panel of Fig. 3, the normalized magnetoresistance, $\Delta
w / w(0)$, is plotted vs $T$ up to 20 K. Quite evidently, phonon
transport emerges as a spectroscopy of magnons. (The same is true
of {\it electron} heat transport, as was recently shown for
CeRhIn$_5$ \cite{Paglione}.) Indeed, the maximum scattering of
phonons will occur when the peak phonon energy matches twice the
average energy of optical magnons, $<E_{magnon}>$ \cite{Dixon}. At
$T=T^{\star}$, heat transport is dominated by thermal phonons with
frequencies around $\omega_{phonon}^{\star} = 4 k_B T^{\star} /
\hbar$. Therefore, in Nd$_2$CuO$_4$, where $<E_{magnon}>$ = 0.5
meV is the average of the measured optical magnon energies (which
range from 0.2 to 0.8 meV \cite{Henggeler,Casalta}), $w(T)$ is
expected to peak when $\hbar \omega_{phonon}^{\star} = 2
<E_{magnon}> \simeq 4 k_B T^{\star}$, {\it i.e.} $ k_BT^{\star}$ =
0.25 meV, or $T^{\star} \simeq 3$~K, precisely as observed.

b) {\underline {Magnons ($T >$ 0.5 K)}}. We assume that magnons in
10 T // $c$ scatter phonons just as well as in 10 T $\perp c$, so
that $\kappa_p$(10 T
// $c$) $\simeq \kappa_p$(10 T $\perp c$), and $\kappa_m$(10 T
$\perp c$) $\simeq$ $\kappa$(10 T $\perp c$) - $\kappa$(10 T
// $c$) $\equiv$ $\Delta\kappa$' (which is equal to $\Delta\kappa$
below 1 K). In the bottom panel of Fig. 3, we plot $\Delta\kappa$'
vs $T$. The magnon conductivity $\kappa_m$ peaks at 3 K, because
the magnon heat capacity $C_m(T)$ saturates when $T$ exceeds the
maximum energy of the acoustic magnon branch, while the inelastic
scattering due to phonons keeps increasing. That maximum energy is
unknown, but we expect it to lie roughly in the range 0.2-0.8 meV.
We estimate $C_m(T)$ using an Einstein model: $C_m$ = $C_E$ =
$z^2$ e$^z$ (e$^z$ -1)$^{-2}$ where $z = \hbar\omega_E/k_BT$, with
$\hbar\omega_E = 0.5$ meV. This is clearly inaccurate at very low
$T$ where $C_m \propto T^3$, but should be reasonable for
temperatures comparable to and above the maximum energy, given
roughly by $\hbar\omega_E$. As seen in Fig. 3, $C_m$ is constant
above 5 K or so and the temperature dependence of $\kappa_m$ comes
entirely from the scattering rate, which we model simply as the
sum of boundary and phonon scattering: $\Gamma = \Gamma_0 +
\Gamma_p$. As seen in Fig. 3, a good fit to $\kappa_m(T)$ is
obtained with $\Gamma_p \propto T^3$ (and $\Gamma_0$ constant, as
it should). This shows that acoustic phonons scatter acoustic
magnons (as they do electrons) in proportion to their number,
which increases as $T^3$. The magnon mean free path in
Nd$_2$CuO$_4$ decreases by a factor of 15 between 5 and 20 K, and
should continue to decrease up to roughly the Debye temperature.
This suggests that the magnon mean free path is unlikely to ever
be independent of temperature in the range 20 to 200 K, as was
assumed in the analysis of data on La$_2$CuO$_4$ \cite{Hess2}.

In summary, by cooling to 50 mK, the characteristic $T^3$ heat
transport expected of ballistic magnons has for the first time
been observed, revealing the existence of an acoustic magnon
branch (gapless or with a very small gap, with upper bound $\sim$
0.1 K) in the collinear phase of Nd$_2$CuO$_4$ ($H >$ 7 T $\perp
c$). Paradoxically, the properties of this mode agree well with
recent calculations applied to the {\it non-collinear} phase ($H$
= 0) \cite{Sachidanandam}, where our data (in $H$ = 0 or $H // c$)
rules it out. A calculation specifically for the collinear phase
in field is desirable, as well as a search for the acoustic branch
via low-energy neutron scattering. Above the ballistic regime,
phonons scatter magnons strongly, initially at a rate $\Gamma_p
\propto T^3$, so that the magnon mean free path decreases rapidly
with temperature, and should continue to do so up to roughly the
Debye temperature. Phonons are themselves strongly scattered by
the magnons, in the temperature range where one phonon-two magnon
processes are most effective, namely in the vicinity of $4k_BT = 2
<E_{magnon}>$. In general, one can expect phonons in all cuprates
(at any doping) to  be scattered significantly by magnetic
excitations (typically from Cu moments).

We are grateful to P. Fournier for useful discussions and sample
characterization, and to I. Affleck, W. Buyers, P. Bourges, C.
Stock, M. Greven, P. Dai, M. Walker for helpful comments. This
research was supported by NSERC of Canada, a Canada Research Chair
(L.T.), and the Canadian Institute for Advanced Research. The work
in China was supported by a grant from the Natural Science
Foundation of China.\\

$^*$ E-mail: louis.taillefer@usherbrooke.ca


\begin{thebibliography}{99}
\bibitem{note} The magnitude of the in-plane magnetic field needed to induce the spin-flop
transition is somewhat anisotropic \cite{Jin}. For our sample,
$\kappa$ starts to increase with field above 1.5 T at 0.3 K, and
saturates beyond 7 T. A field of 10 T was chosen to ensure that
the sample has fully reached the collinear state at all relevant
temperatures.
\bibitem{vp} Note that in Nd$_2$CuO$_4$ $<\hspace{-3pt}v_m^{-2}\hspace{-3pt}>$ is a
3D average over a strongly anisotropic spectrum.
\end{thebibliography}
\end{document}